\documentclass[a4paper,12pt]{iopart}
\usepackage{iopams}
\usepackage{setstack}
\usepackage{graphicx}
\usepackage{bm}
\usepackage{epsfig}

\newcommand{\diag}{{\rm diag\,}}
\newcommand{\sign}{{\rm sign\,}}

\newcommand{\Pf}{{\rm Pf\,}}

\newcommand{\U}{{\rm U\,}}

\newcommand{\Herm}{{\rm Herm\,}}
\newcommand{\Ber}{{\rm Ber\,}}

\newcommand{\eins}{\leavevmode\hbox{\small1\kern-3.8pt\normalsize1}}
\eqnobysec

\begin{document}

\newtheorem{definition}{Definition}[section]
\newtheorem{assumption}[definition]{Assumption}
\newtheorem{theorem}[definition]{Theorem}
\newtheorem{lemma}[definition]{Lemma}
\newtheorem{corollary}[definition]{Corollary}

\title[Pfaffian structures]{A new approach to derive Pfaffian structures for random matrix ensembles}
\author{Mario Kieburg$^\dagger$ and Thomas Guhr}
\address{Universit\"at Duisburg-Essen, Lotharstra\ss e 1, 47048 Duisburg, Germany}
\eads{$^\dagger$ \mailto{mario.kieburg@uni-due.de}}

\date{\today}

\begin{abstract}
 Correlation functions for matrix ensembles with orthogonal and unitary-symplectic rotation symmetry are more complicated to calculate than in the unitary case. The supersymmetry method and the orthogonal polynomials are two techniques to tackle this task. Recently, we presented a new method to average ratios of characteristic polynomials over matrix ensembles invariant under the unitary group. Here, we extend this approach to ensembles with orthogonal and unitary-symplectic rotation symmetry. We show that Pfaffian structures can be derived for a wide class of orthogonal and unitary-symplectic rotation invariant ensembles in a unifying way. This includes also those for which this structure was not known previously, as the real Ginibre ensemble and the Gaussian real chiral ensemble with two independent matrices as well.
\end{abstract}


\section{Introduction}\label{sec1}

There are many applications of Random Matrix Theory in physics as well as in mathematics \cite{Efe97,GMW98,VerWet00,KeaSna00a,Meh04}. Different matrix ensembles describe universal features of eigenvalue statistics in spectra stemming of various physical systems. For example the chiral (Laguerre) ensembles with two independent matrices \cite{Ake02,Osb04,Ake05,APS09} and the Ginibre ensembles \cite{Ede97,Ake01,Ber04,AkeVer03,AkeBas07,AkeKan07,ForNag07} describe universal properties of the Dirac operator and the Hamilton operator with chemical potential.

To model physical systems, one has to use Hermitian matrices, if there are no further constraints. In the case of time reversal invariance, the matrices to be employed are real-symmetric or quaternionic self-dual depending on the behavior of the system under space rotations. These three symmetry classes are referred to as unitary, orthogonal and unitary-symplectic, respectively. Mean values of characteristic polynomial ratios are important quantities to characterize those ensembles. The matrix Green function \cite{Zir06, BorStr05}, the replica trick \cite{VerZir85} as well as the investigation of the sign problem in Quantum Chromodynamics (QCD) \cite{BloWet09} are based on such correlation functions. For many ensembles with factorizing probability density it is known that those averages have a Pfaffian structure \cite{Meh04,BorStr05,Ake05,AkeBas07}. The kernels of the Paffians are mean values of one and two characteristic polynomials. Thus, all eigenvalue correlations are completely determined by the correlations of the lowest order.

The method of orthogonal polynomials \cite{NagFor93,Kan02,ForMay09} and the supersymmetry method \cite{GGK04,SomWie08,APS09} are successful techniques to derive those Pfaffian structures. Recently, we presented a new approach to derive determinantal structures for unitary rotation invariant matrix ensembles in a unifying way \cite{KieGuh09}. Here, we generalize this approach to ensembles with orthogonal and unitary-symplectic rotation symmetry. This also includes real Ginibre ensembles and Gaussian real chiral ensemble with two independent matrices for those the Pfaffian structures  were unknown up to now.

We structure this contribution as follows. In Sec.~\ref{sec2}, we  outline our approach. The general result is presented in Sec.~\ref{sec3}. We give explicit formulae for the real symmetric and the Hermitian self-adjoint matrix ensemble in Sec.~\ref{sec4}. In the same section, we also present two lists of matrix ensembles with orthogonal and unitary-symplectic symmetry to which our method can be applied. Concluding remarks are made in Sec.~\ref{sec6}. In \ref{app1}, we explicitly derive some of the equations in Sec.~\ref{sec3}.

\section{Outline}\label{sec2}

We consider averages of ratios of characteristic polynomials over the Hermitian self-dual matrices
\begin{equation}\label{2.1}
 \displaystyle Z(\kappa)=\int P(H)\prod\limits_{j=1}^k\frac{\det(H-\kappa_{j2}\eins_{2N})}{\det(H-\kappa_{j1}\eins_{2N})}d[H]\,.
\end{equation}
The probability density $P$ is rotation invariant and factorizes in the eigenvalues of $H$, that is, in $E=\diag(E_1,\ldots,E_N)\otimes\eins_2$. We choose $\kappa=\diag(\kappa_{11},\ldots,\kappa_{k1},\kappa_{12},\ldots,\kappa_{k2})=\diag(\kappa_1,\kappa_2)$ in such a way that the integrals are well defined. The matrix $\eins_{2N}$ is the $2N$ dimensional unit matrix. Changing to eigenvalue-angle-coordinates yields
\begin{equation}\label{2.2}
 \displaystyle Z(\kappa)=c\int \prod\limits_{a=1}^N\prod\limits_{b=1}^kP(E_a)\frac{(E_a-\kappa_{b2})^2}{(E_a-\kappa_{b1})^2}\Delta_N^4(E)d[E]
\end{equation}
with a normalization constant $c$. The Vandermonde determinant is defined by
\begin{equation}\label{2.3}
 \displaystyle \Delta_N(E)=\prod\limits_{1\leq a< b\leq N}(E_a-E_b)=(-1)^{N(N-1)/2}\det\left[E_a^{b-1}\right]_{1\leq a,b\leq N}\,.
\end{equation}
Introducing Dirac distributions, we extend the $N$ eigenvalue integrals to $2N$ eigenvalue integrals and have
\begin{equation}\label{2.4}
 \displaystyle Z(\kappa)=c\int\prod\limits_{j=1}^{N} g(E_j,E_{j+N})\prod\limits_{a=1}^{2N}\prod\limits_{b=1}^k\frac{(E_a-\kappa_{b2})}{(E_a-\kappa_{b1})}\Delta_{2N}(E)d[E]\,,
\end{equation}
where
\begin{equation}\label{2.5}
 \displaystyle g(E_j,E_{j+N})=P(E_j)\frac{\delta(E_j-E_{j+N})}{E_j-E_{j+N}}\,.
\end{equation}
In the next step we use the method developed in Ref.~\cite{KieGuh09} based on the idea of Basor and Forrester \cite{BasFor94}. We extend the product of the characteristic polynomials times the Vandermonde determinant by the Cauchy determinant
\begin{equation}\label{2.6}
 \fl\displaystyle\sqrt{\Ber_{(k/k)}^{(2)}(\kappa)}=\frac{\Delta_k(\kappa_1)\Delta_k(\kappa_2)}{\prod\limits_{a,b=1}^k(\kappa_{a1}-\kappa_{b2})}=(-1)^{k(k-1)/2}\det\left[\frac{1}{\kappa_{a1}-\kappa_{b2}}\right]_{1\leq a,b\leq k}\,.
\end{equation}
Notice that the product
\begin{equation}\label{2.6b}
 \Delta_k(\kappa_2)\prod\limits_{a=1}^{2N}\prod\limits_{b=1}^k(E_a-\kappa_{b2})\Delta_{2N}(E)=\Delta_{k+2N}(\kappa_2,E)
\end{equation}
yields another Vandermonde determinant. Then we have
\begin{eqnarray}
 \sqrt{\Ber_{(k/k+2N)}^{(2)}(\kappa_1;\kappa_2,E)}&=&\displaystyle\pm\frac{\Delta_k(\kappa_1)\Delta_{k+2N}(\kappa_2,E)}{\prod\limits_{a,b=1}^k(\kappa_{a1}-\kappa_{b2})\prod\limits_{a=1}^k\prod\limits_{b=1}^N(\kappa_{a1}-E_{b})}=\nonumber\\
 &=&\pm\det\left[\begin{array}{c|c} \displaystyle\underset{\ }{\frac{1}{\kappa_{a1}-\kappa_{b2}}} & \displaystyle\frac{1}{\kappa_{a1}-E_b} \\ \hline \overset{\ }{\kappa_{b2}^{a-1}} & E_b^{a-1} \end{array}\right]\label{2.7}
\end{eqnarray}
times the product of $g$ in the integrand~\eref{2.4}. The notation follows the one in Ref.~\cite{KieGuh09} where we have also proven the second equality in Eq.~\eref{2.7}. A Berezinian is a Jacobian in superspace. The index $(2)$ at the Berezinian refers to Dyson index $\beta=2$. The function $\Ber_{(k/k+2N)}^{(2)}$ appears by diagonalizing Hermitian supermatrices $\sigma=UsU^\dagger$ with a unitary supermatrix $U\in\U(k/k+2N)$, i.e. it is defined by
\begin{eqnarray}
\int f(\sigma)d[\sigma]=\int f(UsU^\dagger) |\Ber^{(2)}_{(k/k+2N)}(s)|d[s]d\mu(U)+{\rm b.t.}\label{2.7b}
\end{eqnarray}
for an arbitrary superfunction $f$. The measure $d\mu(U)$ is the Haar measure on $\U(k/k+2N)$ and ``${\rm b.t.}$'' denotes the Efetov-Wegner boundary terms \cite{Efe83,Weg83,KKG08} which may arise in such a transformation.

As already pointed out in Ref.~\cite{KieGuh09}, this intimate relation to supersymmetry allows us to refer to our approach as ``supersymmetry without supersymmetry'', because we never actually map the matrix model onto superspace. Nevertheless, we establish the previously unknown connection to superspace.

Integrating over all energies $E_j$ with $j>N$ in Eq.~\eref{2.4}, we obtain
\begin{equation}\label{2.8}
 \displaystyle\fl Z(\kappa)=\frac{c}{\sqrt{\Ber_{(k/k)}^{(2)}(\kappa)}}\int\det\left[\begin{array}{c|c|c} \displaystyle\underset{\ }{\frac{1}{\kappa_{a1}-\kappa_{b2}}} & \displaystyle\frac{1}{\kappa_{a1}-E_b} & \displaystyle\int\frac{g(E_b,E)}{\kappa_{a1}-E}dE \\ \hline \overset{\ }{\kappa_{b2}^{a-1}} & E_b^{a-1} & \int g(E_b,E)E^{a-1}dE \end{array}\right]d[E]\,.
\end{equation}
With help of a modified version of de Bruijn's integral theorem \cite{Bru55}, cf. Appendix C.2 in Ref.~\cite{KieGuh09}, we intergrate over the remaining variables and find the Pfaffian expression
\begin{equation}\label{2.9}
 \displaystyle\fl Z(\kappa)=\frac{c}{\sqrt{\Ber_{(k/k)}^{(2)}(\kappa)}}\Pf\left[\begin{array}{c|c|c} 0 &\displaystyle\underset{\ }{\frac{1}{\kappa_{b1}-\kappa_{a2}}} & \kappa_{a2}^{b-1} \\ \hline \displaystyle\overset{\ }{\underset{\ }{\frac{1}{\kappa_{b2}-\kappa_{a1}}}} & \mathbf{F}(\kappa_{a1},\kappa_{b1}) & G_{b}(\kappa_{a1}) \\ \hline -\overset{\ }{\underset{\ }{\kappa_{b2}^{a-1}}} & -G_{a}(\kappa_{b1}) & M_{ab} \end{array}\right]\,.
\end{equation}
We give a detailed definition of the functions $\mathbf{F}$, $G_a$ and $M_{ab}$ in Sec.~\ref{sec3}. Here, we schematically explain what these functions are. The function $\mathbf{F}$ is almost the average over two dimensional Hermitian self-dual matrices of two characteristic polynomials in the denominator. The functions $G_a$ are Cauchy-transforms of the moments of $P$ and $M_{ab}$ is the anti-symmetric moment matrix of $P$ generating the skew orthogonal polynomials of quaternion type \cite{Meh04}.

Since the Pfaffian determinant is skew symmetric in the pairs of rows and columns, we can construct any linear independent set of polynomials in the last columns and rows in Eq.~\eref{2.9}. For example, the skew orthogonal polynomials yield a block diagonal moment matrix $M_{ab}$ which leads immediately to the well known result expressed in terms of skew orthogonal polynomials. Here, we leave the monomials as they are and use
\begin{equation}\label{2.10}
 \Pf\left[\begin{array}{cc} A & B \\ -B^T & D \end{array}\right]=\Pf D\, \Pf[A+BD^{-1}B^T]
\end{equation}
for arbitrary matrices $A$, $B$ and an invertible, anti-symmetric, even dimensional matrix $D$. The matrix $A$ has to be even dimensional and anti-symmetric, too. As $M_{ab}$ is even dimensional, we arrive at the final result
\begin{eqnarray}
 \displaystyle\fl&& Z(\kappa)=\frac{c}{\sqrt{\Ber_{(k/k)}^{(2)}(\kappa)}}\times\nonumber\\
 \fl&\times&\hspace*{-0.1cm}\Pf\hspace*{-0.1cm}\left[\begin{array}{c|c} \displaystyle\sum\limits_{m,n=1}^{2N}\kappa_{a2}^{m-1}M_{mn}^{-1}\kappa_{b2}^{n-1} &\displaystyle\underset{\ }{\frac{1}{\kappa_{b1}-\kappa_{a2}}+\sum\limits_{m,n=1}^{2N}\kappa_{a2}^{m-1}M_{mn}^{-1}G_n(\kappa_{b1})} \\ \hline \displaystyle\overset{\ }{\underset{\ }{\frac{1}{\kappa_{b2}-\kappa_{a1}}+\hspace*{-0.2cm}\sum\limits_{m,n=1}^{2N}\hspace*{-0.1cm}G_m(\kappa_{a1})M_{mn}^{-1}\kappa_{b2}^{n-1}}} & \displaystyle \mathbf{F}(\kappa_{a1},\kappa_{b1})+\hspace*{-0.2cm}\sum\limits_{m,n=1}^{2N}\hspace*{-0.1cm}G_m(\kappa_{a1})M_{mn}^{-1}G_n(\kappa_{b1}) \end{array}\right]\hspace*{-0.1cm}=\nonumber\\
 \fl&=&\frac{c}{\sqrt{\Ber_{(k/k)}^{(2)}(\kappa)}}\Pf\left[\begin{array}{c|c} K_{11}(\kappa_{b2},\kappa_{a2}) & K_{12}(\kappa_{a2},\kappa_{b1}) \\ \hline -K_{12}(\kappa_{a1},\kappa_{b2}) & K_{22}(\kappa_{a1},\kappa_{b1}) \end{array}\right]\,.\label{2.11}
\end{eqnarray}
This is, indeed, the correct result which we found without making use of the Dyson-Mehta-Mahoux theorem \cite{MehMah91}. Although we can employ an arbitrary choice of polynomial set, we obtain the skew orthogonal polynomials generated by $M_{ab}$. Thus, the skew orthogonal polynomials are a result, not an input.

We show in the ensuing sections how Pfaffian structures for a wide class of matrix ensembles can be obtained in a unifying way. Our method is applicable not only for unitary-symplectic symmetry but also for orthogonal symmetric ensembles. We will argue that there is no difference between both symmetries in the derivation. Hence, the Pfaffian structure of averages similar to Eq.~\eref{2.1} is elementary.

\section{Main result}\label{sec3}

We consider the integral
\begin{equation}
  \fl Z_{(k_1/k_2)}^{(2N+1)}(\kappa)=\displaystyle\int\limits_{\mathbb{C}^{2N+1}}h(z_{2N+1})\prod\limits_{j=1}^{N}g(z_{2j-1},z_{2j})\displaystyle\frac{\prod\limits_{a=1}^{2N+1}\prod\limits_{b=1}^{k_2}(z_a-\kappa_{b2})}{\prod\limits_{a=1}^{2N+1}\prod\limits_{b=1}^{k_1}(\kappa_{b1}-z_a)}\Delta_{2N+1}(z)d[z]\,.\label{3.1}
\end{equation}
We choose the functions $h$ and $g$ and the external variables $\kappa=\diag(\kappa_{11},\ldots,\kappa_{k_11},\kappa_{12},$ $\ldots,\kappa_{k_22})$ in such a way that the integral exists. With the two--dimensional Dirac distribution $h(z_{2N+1})=\delta^{(2)}(z_{2N+1})$ and with the function $\tilde{g}(z_{2j-1},z_{2j})=z_{2j-1}z_{2j}g(z_{2j-1},z_{2j})$, we regain another important integral
\begin{equation}\label{3.2}
 \fl\displaystyle Z_{(k_1/k_2)}^{(2N)}(\kappa)=\int\limits_{\mathbb{C}^{2N}}\prod\limits_{j=1}^{N}\tilde{g}(z_{2j-1},z_{2j})\frac{\prod\limits_{a=1}^{k_2}\prod\limits_{b=1}^{2N}(\kappa_{a2}-z_b)}{\prod\limits_{a=1}^{k_1}\prod\limits_{b=1}^{2N}(\kappa_{a1}-z_b)}\Delta_{2N}(z)d[z]\ ,
\end{equation}
which we calculate in the following.

As in Ref.~\cite{KieGuh09}, we extend the integrand in Eq.~\eref{3.1} by $\sqrt{\Ber_{(k_1/k_2)}^{(2)}(\kappa)}$ and obtain
\begin{equation}
 \fl\displaystyle Z_{(k_1/k_2)}^{(2N+1)}(\kappa)=\int\limits_{\mathbb{C}^{2N+1}}h(z_{2N+1})\prod\limits_{j=1}^{N}g(z_{2j-1},z_{2j})\frac{\sqrt{\Ber_{(k_1/k_2+2N+1)}^{(2)}(\tilde{z})}}{\sqrt{\Ber_{(k_1/k_2)}^{(2)}(\kappa)}}d[z]\,,\label{3.3}
\end{equation}
where we define $\tilde{z}=\diag(\kappa_1;\kappa_2,z)$. The extension was made in the same way as described in the previous section. We then use the determinantal structure~\eref{2.7} of the square root Berezinian in the numerator for the integration. In \ref{app1.1} we explicitly calculate Eq.~\eref{3.3} for odd $d=k_2-k_1+2N+1\geq0$ using the sketch of Sec.~\ref{sec2} and find
\begin{eqnarray}
 \fl\displaystyle Z_{(k_1/k_2)}^{(2N+1)}(\kappa)&=&\frac{(-1)^{N+1}N!\Pf\mathbf{M}_{(d)}}{\sqrt{\Ber_{(k_1/k_2)}^{(2)}(\kappa)}}\times\nonumber\\
 \fl&\times&\Pf\left[\begin{array}{cc} \left\{K_{11}^{(d)}(\kappa_{a2},\kappa_{b2})\right\}\underset{1\leq a,b\leq k_2}{\ } & \left\{K_{12}^{(d)}(\kappa_{b1},\kappa_{a2})\right\}\underset{1\leq b\leq k_1}{\underset{1\leq a\leq k_2}{\ }} \\ \left\{-K_{12}^{(d)}(\kappa_{a1},\kappa_{b2})\right\}\underset{1\leq b\leq k_2}{\underset{1\leq a\leq k_1}{\ }} & \left\{K_{22}^{(d)}(\kappa_{a1},\kappa_{b1})\right\}\underset{1\leq a,b\leq k_1}{\ } \end{array}\right]\,,\label{3.4}
\end{eqnarray}
where
\begin{eqnarray}
 \fl \mathbf{F}(\kappa_{a1},\kappa_{b1})&=&-(\kappa_{a1}-\kappa_{b1})Z_{(2/0)}^{(2)}(\kappa_{a1},\kappa_{b1})=\nonumber\\
 \fl&=&-(\kappa_{a1}-\kappa_{b1})\int\limits_{\mathbb{C}^2}\frac{g(z_1,z_2)(z_1-z_2)}{(\kappa_{a1}-z_1)(\kappa_{a1}-z_2)(\kappa_{b1}-z_1)(\kappa_{b1}-z_2)} d[z]\label{3.5}\,,\\
 \fl\mathbf{G}_{(d)}(\kappa_{a1})&=&\left[\begin{array}{cc} \left\{\displaystyle\int\limits_{\mathbb{C}^2}\hspace*{-0.15cm}\det\left[\begin{array}{cc} \displaystyle\frac{g(z_1,z_2)}{\kappa_{a1}-z_1} & \displaystyle\frac{g(z_1,z_2)}{\kappa_{a1}-z_2} \\ \displaystyle z_1^{b-1} & \displaystyle z_2^{b-1} \end{array}\right]d[z]\right\}_{1\leq b\leq d} & \displaystyle-\hspace*{-0.15cm}\int\limits_{\mathbb{C}}\hspace*{-0.15cm}\frac{h(z)}{\kappa_{a1}-z}dz \end{array}\right]\label{3.6},\\
 \fl\mathbf{K}_{(d)}(\kappa_{a2})&=&\left[\begin{array}{cc} \left\{\kappa_{a2}^{b-1}\right\}_{1\leq b\leq d} & 0 \end{array}\right]\label{3.7}\,,\\
 \fl K_{11}^{(d)}(\kappa_{a2},\kappa_{b2})&=&\mathbf{K}_{(d)}(\kappa_{a2})\mathbf{M}_{(d)}^{-1}\mathbf{K}_{(d)}^T(\kappa_{b2})\label{3.8}\,,\\
 \fl K_{12}^{(d)}(\kappa_{b1},\kappa_{a2})&=&\displaystyle\frac{1}{\kappa_{b1}-\kappa_{a2}}+\mathbf{K}_{(d)}(\kappa_{a2})\mathbf{M}_{(d)}^{-1}\mathbf{G}_{(d)}^T(\kappa_{b1})\label{3.9}\,,\\
 \fl K_{22}^{(d)}(\kappa_{a1},\kappa_{b1})&=&\mathbf{F}(\kappa_{a1},\kappa_{b1})+\mathbf{G}_{(d)}(\kappa_{a1})\mathbf{M}_{(d)}^{-1}\mathbf{G}_{(d)}^T(\kappa_{b1})\label{3.10}\,.
\end{eqnarray}
Here, we use the moment matrix
\begin{equation}
 \fl\mathbf{M}_{(d)}\hspace*{-0.15cm}=\hspace*{-0.15cm}\left[\begin{array}{cc} \left\{\displaystyle\int\limits_{\mathbb{C}^2}\hspace*{-0.15cm}\det\left[\begin{array}{cc} g(z_1,z_2)z_1^{a-1} & z_1^{b-1} \\ g(z_1,z_2)z_2^{a-1} & z_2^{b-1} \end{array}\right]d[z]\right\}_{1\leq a,b\leq d} & \left\{\displaystyle-\hspace*{-0.15cm}\int\limits_{\mathbb{C}}\hspace*{-0.15cm}h(z)z^{a-1}dz\right\}_{1\leq a\leq d} \\ \left\{\displaystyle\int\limits_{\mathbb{C}}h(z)z^{b-1}dz\right\}_{1\leq b\leq d} & 0 \end{array}\right]\label{3.11}
\end{equation}
of our probability densities $h$ and $g$. Let $\mathfrak{S}_M$ be the permutation group of $M$ elements and the function ``$\sign$'' equals ``$+1$'' for even permutations and ``$-1$'' for odd ones. We fix the sign of the Pfaffian for an arbitrary anti-symmetric $2N\times2N$ matrix $\{D_{ab}\}$ by
\begin{equation}\label{3.0}
 \Pf[D_{ab}]_{1\leq a,b\leq N}=\frac{1}{2^NN!}\sum\limits_{\omega\in\mathfrak{S}_{2N}}\sign(\omega)\prod\limits_{j=1}^ND_{\omega(2j)\omega(2j+1)}\,.
\end{equation}
The sums in Eq.~\eref{2.11} for Hermitian self-dual matrices are encoded in the matrix products of Eqs.~\eref{3.8}, \eref{3.9} and \eref{3.10}. Indeed, we see that the sketch of the approach in Sec.~\ref{sec2} can readily be extended from this particular case to the quite general integral~\eref{3.1}.

The integral kernels \eref{3.8} to \eref{3.10} have a simple relation to the generating function~\eref{3.1} with other parameters than $k_1$, $k_2$ and $N$. We identify them with the particular cases ($k_1=0,\,k_2=2$), ($k_1=1,\,k_2=1$) and ($k_1=2,\,k_2=0$),
\begin{eqnarray}
 \fl K_{11}^{(2N+3)}(\kappa_{a2},\kappa_{b2})&=&\displaystyle(-1)^{N+1}\frac{\kappa_{a2}-\kappa_{b2}}{N!\Pf\mathbf{M}_{(2N+3)}} Z_{(0/2)}^{(2N+1)}(\kappa_{a2},\kappa_{b2})\label{3.12}\,,\\
 \fl K_{12}^{(2N+1)}(\kappa_{b1},\kappa_{a2})&=&\displaystyle(-1)^{N+1}\frac{1}{N!\Pf\mathbf{M}_{(2N+1)}(\kappa_{b1}-\kappa_{a2})} Z_{(1/1)}^{(2N+1)}(\kappa_{b1},\kappa_{a2})\label{3.13}\,,\\
 \fl K_{22}^{(2N-1)}(\kappa_{a1},\kappa_{b1})&=&\displaystyle(-1)^{N+1}\frac{\kappa_{a1}-\kappa_{b1}}{N!\Pf\mathbf{M}_{(2N-1)}} Z_{(2/0)}^{(2N+1)}(\kappa_{a1},\kappa_{b1})\label{3.14}\,.
\end{eqnarray}
The normalization constant is defined by the case $k_1=k_2=0$,
\begin{equation}\label{3.15}
 C_{(2N+1)}=Z_{(0/0)}^{(2N+1)}=(-1)^{N+1}N!\Pf\mathbf{M}_{(2N+1)}\,.
\end{equation}
Hence, we plug these relations into Eq.~\eref{3.4} which leads to the result
\begin{eqnarray}
 \fl&&\displaystyle Z_{(k_1/k_2)}^{(2N+1)}(\kappa)=\frac{(-1)^{(k_2^2-k_1^2)/4+k_1+1}N!\left[(-1)^{N}\Pf\mathbf{M}_{(d)}\right]^{1-(k_1+k_2)/2}}{\sqrt{\Ber_{(k_1/k_2)}^{(2)}(\kappa)}}\times\label{3.16}\\
 \fl&\times&\Pf\left[\begin{array}{cc} \left\{\displaystyle\frac{(\kappa_{b2}-\kappa_{a2})Z_{(0/2)}^{(d-2)}(\kappa_{a2},\kappa_{b2})}{[(d-3)/2]!} \right\}\underset{1\leq a,b\leq k_2}{\ } & \left\{\displaystyle\frac{Z_{(1/1)}^{(d)}(\kappa_{b1},\kappa_{a2})}{[(d-1)/2]!(\kappa_{b1}-\kappa_{a2})} \right\}\underset{1\leq b\leq k_1}{\underset{1\leq a\leq k_2}{\ }} \\ \left\{\displaystyle\frac{Z_{(1/1)}^{(d)}(\kappa_{a1},\kappa_{b2})}{[(d-1)/2]!(\kappa_{b2}-\kappa_{a1})}\right\}\underset{1\leq b\leq k_2}{\underset{1\leq a\leq k_1}{\ }} & \left\{\displaystyle\frac{(\kappa_{b1}-\kappa_{a1})Z_{(2/0)}^{(d+2)}(\kappa_{a1},\kappa_{b1})}{[(d+1)/2]!} \right\}\underset{1\leq a,b\leq k_1}{\ } \end{array}\right].\nonumber
\end{eqnarray}
When $d$ is odd, $k_1+k_2$ is even. Thus, the Pfaffians are well defined. Equation~\eref{3.16} implies that the correlations for two characteristic polynomials determine all other eigenvalue correlations if the probability density has the factorizing structure as in Eq.~\eref{3.1}.

For the case that $k_2+k_1$ is odd, we extend the integral
\begin{equation}
  Z_{(k_1/k_2)}^{(2N+1)}(\kappa)=-\displaystyle\underset{\kappa_{02}\to\infty}{\lim}\frac{Z_{(k_1/k_2+1)}^{(2N+1)}(\kappa)}{\kappa_{02}^{2N+1}}\label{3.17}
\end{equation}
by an additional parameter $\kappa_{02}$. This trick is similar to the one in Refs.~\cite{SWG99,KieGuh09}. Defining $\tilde{d}=k_2-k_1+2N+2\geq0$, we apply the result~\eref{3.16} and find
\begin{eqnarray}
 \fl&&\displaystyle Z_{(k_1/k_2)}^{(2N+1)}(\kappa)=\frac{(-1)^{(k_2+k_1+1)/2}N!}{\sqrt{\Ber_{(k_1/k_2)}^{(2)}(\kappa)}}\times\label{3.18}\\
 \fl&\times&\Pf\left[\begin{array}{ccc} 0 & \left\{\displaystyle\frac{-Z_{(0/1)}^{(\tilde{d}-2)}(\kappa_{b2})}{[(\tilde{d}-3)/2]!}\right\}\underset{1\leq b\leq k_2}{\ } & \left\{\displaystyle\frac{-Z_{(1/0)}^{(\tilde{d})}(\kappa_{b1})}{[(\tilde{d}-1)/2]!}\right\}\underset{1\leq b\leq k_1}{\ } \\ \left\{\displaystyle\frac{Z_{(0/1)}^{(\tilde{d}-2)}(\kappa_{a2})}{[(\tilde{d}-3)/2]!}\right\}\underset{1\leq a\leq k_2}{\ } & \left\{K_{11}^{(\tilde{d})}(\kappa_{a2},\kappa_{b2})\right\}\underset{1\leq a,b\leq k_2}{\ } & \left\{K_{12}^{(\tilde{d})}(\kappa_{b1},\kappa_{a2})\right\}\underset{1\leq b\leq k_1}{\underset{1\leq a\leq k_2}{\ }} \\ \left\{\displaystyle\frac{Z_{(1/0)}^{(\tilde{d})}(\kappa_{a1})}{[(\tilde{d}-1)/2]!}\right\}\underset{1\leq a\leq k_1}{\ } & \left\{-K_{12}^{(\tilde{d})}(\kappa_{a1},\kappa_{b2})\right\}\underset{1\leq b\leq k_2}{\underset{1\leq a\leq k_1}{\ }} & \left\{K_{22}^{(\tilde{d})}(\kappa_{a1},\kappa_{b1})\right\}\underset{1\leq a,b\leq k_1}{\ } \end{array}\right]\nonumber
\end{eqnarray}
in the limit $\kappa_{02}\to\infty$. We notice the appearance of one--point functions. Equation~\eref{3.18} is the analog for odd $k_2+k_1$ to the result~\eref{3.16} which is true for even $k_2+k_1$.

The results~\eref{3.16} and \eref{3.18} are also true for the integral \eref{3.2}. We simply have to choose $h$ as a Dirac distribution. This relation is well known \cite{Meh04} for odd and even dimensional ensembles over real symmetric matrices or circular orthogonal matrices. Since the probability densities $g$ and $h$ are quite arbitrary this result considerably extends the one found by Borodin and Strahov \cite{BorStr05}.

We are also interested in the case of $d=k_2-k_1+2N+1\leq0$. Employing the sketched derivation in \ref{app1.2}, we have the result
\begin{eqnarray}
 \fl&&\displaystyle Z_{(k_1/k_2)}^{(2N+1)}(\kappa)=\frac{(-1)^{N}N!}{\sqrt{\Ber_{(k_1/k_2)}^{(2)}(\kappa)}}\times\label{3.19}\\
 \fl&\times&\hspace*{-0.05cm}\displaystyle\Pf\hspace*{-0.15cm}\left[\begin{array}{cccc} 0 & 0 & 0 & \hspace*{-0.3cm}\left\{\displaystyle\frac{1}{\kappa_{b1}-\kappa_{a2}}\right\}\underset{1\leq b\leq k_1}{\underset{1\leq a\leq k_2}{\ }} \\ 0 & 0 & 0 & \hspace*{-0.3cm}\left\{\displaystyle Z_{(1/0)}^{(1)}(\kappa_{b1})\right\}_{1\leq b\leq k_1} \\ 0 & 0 & 0 & \hspace*{-0.3cm}\left\{\displaystyle \kappa_{b1}^{a-1}\right\}\underset{1\leq b\leq k_1}{\underset{1\leq a\leq -d}{\ }} \\ \hspace*{-0.15cm}\left\{\displaystyle\frac{1}{\kappa_{b2}-\kappa_{a1}}\right\}\underset{1\leq b\leq k_2}{\underset{1\leq a\leq k_1}{\ }} & \hspace*{-0.35cm}\left\{\displaystyle -Z_{(1/0)}^{(1)}(\kappa_{a1})\right\}_{1\leq a\leq k_1} & \hspace*{-0.35cm}\left\{\displaystyle -\kappa_{a1}^{b-1}\right\}\underset{1\leq b\leq -d}{\underset{1\leq a\leq k_1}{\ }} & \hspace*{-0.35cm}\left\{\mathbf{F}(\kappa_{a1},\kappa_{b1})\right\}_{1\leq a,b \leq k_1}  \end{array}\right]\,.\nonumber
\end{eqnarray}
For the integral~\eref{3.2}, we have to omit the column and the row with $Z_{(1/0)}^{(1)}$ and have to replace $d$ by $2N+k_2-k_1$. The matrix in the Pfaffian~\eref{3.19} is, indeed, even dimensional. Thus, the expression is well defined.

The Pfaffian structure of the sparsely occupied matrix~\eref{3.19} for $d\leq0$ is a new result. The row and the column with $Z_{(1/0)}^{(1)}$ only appears for odd dimensional, real symmetric matrices. This factor is the Cauchy--transform of the probability density itself. The function $\mathbf{F}$ is almost the mean value of the two characteristic polynomials in the denominator which has to be calculated, too. However, the $N$ eigenvalue integrals are drastically reduced to one or two dimensional integrals.

\section{Applications}\label{sec4}

In Sec.~\ref{sec4.1}, we apply the general results to two ensembles over real symmetric matrices and Hermitian self-dual matrices. We give an overview of applications for ensembles which are rotation invariant under the orthogonal and unitary-symplectic group in Sec.~\ref{sec4.2}.

\subsection{Rotation invariant ensembles over real symmetric matrices and Hermitian self-dual matrices}\label{sec4.1}

We consider mean values of characteristic polynomials for a rotation invariant probability density $P$ over the real symmetric matrices $\Herm(1,N)$ or the Hermitian self-adjoint matrices $\Herm(4,N)$, respectively,
\begin{equation}\label{4.1}
 \displaystyle Z_{(k_1/k_2)}^{(N,\beta)}(\kappa)=\int\limits_{\Herm(\beta,N)}P(H)\frac{\prod\limits_{j=1}^{k_2}\det(H-\kappa_{j2}\eins_{\gamma N})}{\prod\limits_{j=1}^{k_1}\det(H-\kappa_{j1}\eins_{\gamma N})}d[H]\,.
\end{equation}
The constant $\gamma$ equals one for the real case and two for the quaternionic case. Such averages were considered before \cite{LehSom91,GGK04,BorStr05,KGG08}. Here, we apply our method to show that the Pfaffian structure arises in a purely algebraic way. As far as we know this is a new insight.

The generating function~\eref{4.1} is related to the matrix Green function and thus to the $k$-point correlation functions. These matrix ensembles describe time-reversal invariant systems.

For the quaternionic case, the diagonalization of $H$ leads to the identification
\begin{equation}\label{4.2}
 \tilde{g}(z_1,z_2)=P(x_{1})\delta(y_{1})\delta(y_{2})\frac{\delta(x_{2}-x_{1})}{x_1-x_2}\,,
\end{equation}
c.f. Eq.~\eref{3.2}, and
\begin{equation}\label{4.3}
 Z_{(k_1/k_2)}^{(2N)}(\kappa)=\frac{N!}{C_N^{(4)}}Z_{(k_1/k_2)}^{(N,4)}(\kappa)
\end{equation}
with
\begin{equation}\label{4.3b}
C_N^{(4)}=(-1)^{N(N-1)/2}\prod\limits_{j=1}^N\frac{\pi^{2(j-1)}}{\Gamma(2j)}\,.
\end{equation}
Hence, we plug these relations into Eq.~\eref{3.16} for the case $c=N+(k_2-k_1)/2\in\mathbb{N}$ and find
\begin{eqnarray}
 \fl&&\displaystyle Z_{(k_1/k_2)}^{(N,4)}(\kappa)=\frac{C_N^{(4)}Z_{(0/0)}^{(c,4)}}{C_c^{(4)}\sqrt{\Ber_{(k_1/k_2)}^{(2)}(\kappa)}}\times\label{4.3c}\\
 \fl&\times&\Pf\left[\begin{array}{c|c} \displaystyle\underset{\ }{(\kappa_{b2}-\kappa_{a2})\frac{C_{c}^{(4)}Z_{(0/2)}^{(c-1,4)}(\kappa_{a2},\kappa_{b2})}{C_{c-1}^{(4)}Z_{(0/0)}^{(c,4)}}} & \displaystyle\frac{Z_{(1/1)}^{(c,4)}(\kappa_{b1},\kappa_{a2})}{(\kappa_{b1}-\kappa_{a2})Z_{(0/0)}^{(c,4)}} \\ \hline \overset{\ }{\displaystyle\frac{Z_{(1/1)}^{(c,4)}(\kappa_{a1},\kappa_{b2})}{(\kappa_{b2}-\kappa_{a1})Z_{(0/0)}^{(c,4)}}} & \displaystyle(\kappa_{b1}-\kappa_{a1})\frac{C_{c}^{(4)}Z_{(2/0)}^{(c+1,4)}(\kappa_{a1},\kappa_{b1})}{C_{c+1}^{(4)}Z_{(0/0)}^{(c,4)}} \end{array}\right]\,.\nonumber
\end{eqnarray}
The indices $a$ and $b$ numerate all variables $\kappa$ and, thus, the upper left block and the lower right block are a $k_2\times k_2$ matrix and a $k_1\times k_1$ matrix, respectively. Similarly, one finds results for the cases of odd $k_2-k_1$ or negative integer $2N+k_2-k_1+1$ according to the equations~\eref{3.18} and \eref{3.19}.

Let $N=2L+\chi$ with $\chi\in\{0,1\}$. The diagonalization in the real case leads to a product of Heavyside distributions $\Theta(E_{j+1}-E_j)$, $j\in\{1,\ldots,N-1\}$, which is equivalent to the ordering of the eigenvalues $E_1\leq E_2\leq\ldots\leq E_N$. Let $z_j=E_j+\imath y_j$. We split the product of Heavyside distributions in two products
\begin{equation}\label{4.4}
 \prod\limits_{j=1}^{N-1}\Theta(E_{j+1}-E_j)=\prod\limits_{j=1}^{L+\chi-1}\Theta(E_{2j+1}-E_{2j})\prod\limits_{j=1}^{L}\Theta(E_{2j}-E_{2j-1})\,.
\end{equation}
We absorb the second product of Eq.~\eref{4.4} into the probability density and define the probability densities
\begin{equation}\label{4.5}
 g(z_{1},z_{2})=\tilde{g}(z_1,z_2)=P(E_{1})P(E_{2})\delta(y_{1})\delta(y_{2})\Theta(E_{2}-E_{1})
\end{equation}
and
\begin{equation}\label{4.6}
 h(z)=P(E)\delta(y)\,,
\end{equation}
according to even and odd $N$. Due to the integration method over alternate variables \cite{Meh67}, we identify
\begin{equation}\label{4.7}
  Z_{(k_1/k_2)}^{(2L+\chi)}(\kappa)=\frac{L!}{C_{2L+\chi}^{(1)}} Z_{(k_1/k_2)}^{(2L+\chi,1)}(\kappa)
\end{equation}
with
\begin{equation}\label{4.7b}
C_{2L+\chi}^{(1)}=(-1)^{\chi k_1}\prod\limits_{j=1}^{2L+\chi}\frac{\pi^{(j-1)/2}}{\Gamma(j/2)}\,.
\end{equation}
Again, we rewrite Eq.~\eref{3.16} with help of relation~\eref{4.7} and have for even $\tilde{c}=2L+k_2-k_1$
\begin{eqnarray}
 \fl&&\displaystyle Z_{(k_1/k_2)}^{(2L+\chi,1)}(\kappa)=\frac{C_{2L+\chi}^{(1)}Z_{(0/0)}^{(\tilde{c}+\chi,1)}}{C_{\tilde{c}+\chi}^{(1)}\sqrt{\Ber_{(k_1/k_2)}^{(2)}(\kappa)}}\times\label{4.7c}\\
 \fl&\times&\Pf\left[\begin{array}{c|c} \displaystyle\underset{\ }{(\kappa_{b2}-\kappa_{a2})\frac{C_{\tilde{c}+\chi}^{(1)}Z_{(0/2)}^{(\tilde{c}+\chi-2,1)}(\kappa_{a2},\kappa_{b2})}{C_{\tilde{c}+\chi-2}^{(1)}Z_{(0/0)}^{(\tilde{c}+\chi,1)}}} & \displaystyle\frac{Z_{(1/1)}^{(\tilde{c}+\chi,1)}(\kappa_{b1},\kappa_{a2})}{(\kappa_{b1}-\kappa_{a2})Z_{(0/0)}^{(\tilde{c}+\chi,1)}} \\ \hline \overset{\ }{\displaystyle\frac{Z_{(1/1)}^{(\tilde{c}+\chi,1)}(\kappa_{a1},\kappa_{b2})}{(\kappa_{b2}-\kappa_{a1})Z_{(0/0)}^{(\tilde{c}+\chi,1)}}} & \displaystyle(\kappa_{b1}-\kappa_{a1})\frac{C_{\tilde{c}+\chi}^{(1)}Z_{(2/0)}^{(\tilde{c}+\chi+2,1)}(\kappa_{a1},\kappa_{b1})}{C_{\tilde{c}+\chi+2}^{(1)}Z_{(0/0)}^{(\tilde{c}+\chi,1)}} \end{array}\right]\,.\nonumber
\end{eqnarray}
The indices $a$ and $b$ numerate all variables $\kappa$ as in Eq.~\eref{4.3c}. Similar results are valid for the cases of odd $k_2-k_1$ or negative integer $2N+k_2-k_1+1$.

Let $d=k_2-k_1+\gamma N\geq0$. The moment matrices
\begin{equation}\label{4.8}
 \fl M_{(d)}^{(1)}=\left[\underset{-\infty\leq E_1\leq E_2\leq\infty}{\int\hspace*{-0.3cm}\int}P(E_1)P(E_2)(E_1^{a-1}E_2^{b-1}-E_1^{b-1}E_2^{a-1})dE_1dE_2\right]_{1\leq a,b\leq d}
\end{equation}
for the real case with even $d$,
\begin{equation}\label{4.9}
 \widetilde{M}_{(d)}^{(1)}=\left[\begin{array}{cc} M_{(d)}^{(1)} & \left\{-\int\limits_{\mathbb{R}}P(E)E^{a-1}dE\right\}_{1\leq a\leq d} \\ \left\{\int\limits_{\mathbb{R}}P(E)E^{b-1}dE\right\}_{1\leq b\leq d} & 0 \end{array}\right]
\end{equation}
for the real case with odd $d$ and
\begin{equation}\label{4.10}
 M_{(d)}^{(4)}=\left[(a-b)\int\limits_{\mathbb{R}}P(E)E^{a+b-3}dE\right]_{1\leq a,b\leq d}
\end{equation}
for the quaternionic case generate the skew orthogonal polynomials, corresponding to the symmetry. Considering the structure of the Berezinian, this shows an intimate connection between the method of orthogonal polynomials and the supersymmetry method.

The Pfaffian structures~\eref{4.3c} and \eref{4.7c} are well known \cite{BorStr05}. However, those Pfaffian structures involving the insertion of the relations~\eref{4.3} and \eref{4.7} into Eq.~\eref{3.19} are new results. They show that something crucial happens when $2N+k_2-k_1+1$ is negative. The derivation of these structures is purely algebraical. Hence, it is independent of the probability density under consideration. Even with help of the supersymmetry method one could not reduce the number of integrals in such a substantial way.

Another new insight of paramount importance is that the structures obtained for real symmetric matrices and those for Hermitian self-dual matrices have a common origin. In all the other methods \cite{Meh04,GGK04,BorStr05} both cases were considered separately. Our method shows that in both cases, in the real and in the quaternionic one, the underlying algebraic structure yielding Pfaffian determinants is the same.

\subsection{A list of other matrix ensembles}\label{sec4.2}

We average ratios of characteristic polynomials similar to the type~\eref{4.1} where the integration domains are matrix sets different from the symmetric spaces. Those matrix sets have to be rotation invariant either under the orthogonal group or under the unitary symplectic group. For both symmetries we give a list of ensembles to which the integrals~\eref{3.1} or \eref{3.2} are applicable. A real--imaginary part decomposition is $z_j=x_j+\imath y_j$ and in polar coordinates it is $z_j=r_je^{\imath\varphi_j}$. Then, the probability densities in Eqs.~\eref{3.1} and \eref{3.2} are equivalent to the probability densities in Eq.~\eref{4.1} after suitable changes of variables. The ensembles with orthogonal symmetry are given in table~\ref{t1} and those with unitary-symplectic symmetry are listed in table~\ref{t2}.

\begin{table}[tbp] \centering
\rotatebox{90}{
\begin{tabular}[c]{l|c|c|c|c}
 matrix ensemble & probability density $P$ &  matrices in the & probability & probability \\
  & for the matrices & characteristic & densities $g(z_{1},z_{2})$ & density $h(z)$ \\
  & & polynomials & and $\tilde{g}(z_1,z_2)$ &\\
 \noalign{\vskip\doublerulesep\hrule height 2pt}
 real symmetric matri-  & $\overset{}{\widetilde{P}}\left(\tr H^m,m\in\mathbb{N}\right)$ & $H$ & $P(x_{1})P(x_{2})\times$ & $P(x)\delta(y)$ \\
 ces \cite{LehSom91,GGK04,BorStr05} & $H=H^T=H^*$ & & $\times\delta(y_{1})\delta(y_{2})\Theta(x_{2}-x_{1})$ & \\ \hline 
 circular orthogonal & $\overset{}{\widetilde{P}}\left(\tr U^m,m\in\mathbb{N}\right)$ & $U$ and $U^\dagger$ & $P(e^{\imath\varphi_{1}})P(e^{\imath\varphi_{2}})\times$ & $P(e^{\imath\varphi})\delta(r-1)$ \\ 
 ensemble \cite{KeaSna00a} & $U^\dagger U=\eins_N$ and & & $\times\delta(r_{1}-1)\delta(r_{2}-1)\times$ & \\
  & $U^T=U$ & & $\times\Theta(\varphi_{2}-\varphi_{1})$ & \\ \hline
 real symmetric chiral & $\overset{\ }{\widetilde{P}}\left(\tr (AA^T)^m,m\in\mathbb{N}\right)$ & $AA^T$ &  $P(x_1)P(x_2)\times$ & $P(x)\delta(y)x^{(\nu-1)/2}$ \\ 
 (real Laguerre) & $A$ is a real $N\times M$  & & $\times (x_1x_2)^{(\nu-1)/2}\times$ & \\
 ensemble &  matrix with & & $\times\delta(y_{1})\delta(y_{2})\Theta(x_{2}-x_{1})$ & \\
 \cite{NagFor93,For93,Ver94,NagFor95}& $\nu=M-N\geq0$ & & & \\ \hline
 Gaussian real elliptical & $\displaystyle\overset{\ }{\exp\left[-\frac{(\tau+1)}{2}\tr H^T H\right]\times}$ & $\overset{\ }{H}$ & $\displaystyle\prod\limits_{j\in\{1,2\}}\exp\left[-\tau x_j^2\right]\times$ & $\exp(-\tau x^2)\delta(y)$ \\
 ensemble; for $\tau=1$ & $\displaystyle\times\exp\left[-\frac{(\tau-1)}{2}\tr H^2\right]$ & & $\times\sqrt{{\rm erfc}(\sqrt{2(1+\tau)}y_j)}\times$ & \\
 real Ginibre ensemble & $H=H^*$; & & $\times[\delta(y_1)\delta(y_2)\Theta(x_2-x_1)+$ & \\
 \cite{Ede97,AkeKan07,Som07,ForNag07,BorSin08a,SomWie08} & $\tau>0$ & & $+2\imath\delta^{(2)}(z_1-z_2^*)\Theta(y_1)]$ & \\
 \cite{BorSin08b,ForNag08,ForMay09} & & & & \\ \hline
 & $\exp\left[-\tr A^T A-\tr B^T B\right]$ & $CD$ & $\displaystyle\prod\limits_{j\in\{1,2\}}\exp\left[-2\eta_-z_j\right]\times$ & $x^{\nu/2}\exp\left[-2\eta_-x\right]\times$ \\
 Gaussian real chiral & $C=A+\mu B$ &  & $\times |z_j|^\nu\sqrt{f(2\eta_+z_j)}\times$ & $\times K_{\nu/2}(2\eta_+x)\delta(y)$ \\
 ensemble \cite{APS09,APS09b} & $D=-A^T+\mu B^T$ & & $\times[\delta(y_1)\delta(y_2)\Theta(x_2-x_1)+$ & \\
  & $A$ and $B$ are & & $+2\imath\delta^{(2)}(z_1-z_2^*)\Theta(y_1)]$ & \\
 & real $N\times M$ matrices & & & \\
  & with $\nu=M-N\geq0$ & & &
\end{tabular}}
\caption{\label{t1}  Particular cases of the probability densities $g(z_1,z_2)$ and $h(z)$ and their corresponding matrix ensembles of orthogonal rotation symmetry. The joint probability density is equivalent to $g(z_1,z_2)$ and $h(z)$. The density $h(z)$ only appears for odd dimensional matrices.}
\end{table}

\begin{table}[tbp] \centering
\rotatebox{90}{
\begin{tabular}[c]{l|c|c|c}
 matrix ensemble & probability density $P$ &  matrices in the & probability density \\
  & for the matrices & characteristic & $\tilde{g}(z_{1},z_{2})$  \\
  &  & polynomials &  \\
 \noalign{\vskip\doublerulesep\hrule height 2pt}
 Hermitian, self-dual matrices  & $\overset{}{\widetilde{P}}\left(\tr H^m,m\in\mathbb{N}\right)$ & $H$ & $\displaystyle\overset{\ }{P(x_{1})\delta(y_{1})\delta(y_2)\frac{\delta(x_{2}-x_{1})}{x_1-x_2}}$ \\
 \cite{GGK04,BorStr05} & $H=H^\dagger$ & & \\ \hline 
 circular unitary-symplectical & $\overset{}{\widetilde{P}}\left(\tr U^m,m\in\mathbb{N}\right)$ & $U$ and $U^\dagger$ & $P\left(e^{\imath\varphi_1}\right)\delta(r_1-1)\times$ \\ 
  ensemble \cite{KeaSna00a} & $U^\dagger U=\eins_N$ & & $\displaystyle\times\delta(r_2-1)\frac{\delta(\varphi_{2}-\varphi_{1})}{\sin(\varphi_1-\varphi_2)}$ \\
  & & & \\ \hline
 Hermitian self-dual chiral & $\overset{\ }{\widetilde{P}}\left(\tr (AA^\dagger)^m,m\in\mathbb{N}\right)$ & $AA^\dagger$ &  $\displaystyle\overset{\ }{P(x_{1})x_1^{M-N+1}\delta(y_{1})\delta(y_2)\times}$ \\ 
 (quaternionic Laguerre) & $A$ is a quaternionic $N\times M$  & & $\displaystyle\times\frac{\delta(x_{2}-x_{1})}{x_1-x_2}$ \\
 ensemble \cite{NagFor93,For93,Ver94,NagFor95}& matrix with $N\leq M$ & &  \\ \hline
 Gaussian quaternionic ellipti- & $\displaystyle\overset{\ }{\exp\left[-\frac{(\tau+1)}{2}\tr H^T H\right]\times}$ & $\overset{\ }{H}$ & $\exp\left[-2r_1^2(\sin^2\varphi_1+\tau\cos^2\varphi_1)\right]\times$ \\
 cal ensemble; for $\tau=1$ quater- & $\displaystyle\times\exp\left[-\frac{(\tau-1)}{2}\tr H^2\right]$ &  & $\displaystyle\times r_1\sin(2\varphi_1)\delta(r_1-r_2)\delta(\varphi_1+\varphi_2)$ \\
 nionic Ginibre ensemble \cite{Kan02,AkeBas07} & $H$ is a quaternionic matrix & & \\ \hline
 & $\exp\left[-\tr A^\dagger A-\tr B^\dagger B\right]$ & $CD$ & $\displaystyle K_{2\nu}\left(2\eta_+r_1\right)r_1^{2\nu}\times$\\
 Gaussian quaternionic chiral & $C=\imath A+\mu B$ &  & $\displaystyle\times \exp\left[2\eta_-r_1\cos\varphi_1\right]\times$ \\
 ensemble \cite{Ake05} & $D=\imath A^\dagger+\mu B^\dagger$ & & $\times r_1\sin\varphi_1\delta(r_1-r_2)\delta(\varphi_1+\varphi_2)$ \\
 & $A$ and $B$ are quaternionic & & \\
  & $N\times M$ matrices  & &\\
  &  with $\nu=M-N\geq0$ & &
\end{tabular}}
\caption{\label{t2}  Particular cases of the probability densities $\tilde{g}(z_1,z_2)$ and their corresponding matrix ensembles unitary--symplectic rotation symmetry. The joint probability density is equivalent to $\tilde{g}(z_1,z_2)$. All matrices have quaternion structure and, thus, they are even dimensional.}
\end{table}

Our method can be applied to the examples in both tables by the same procedure as described in Sec.~\ref{sec4.1}. One identifies the probability densities $g$, $\tilde{g}$ and $h$ in Eqs.~\eref{3.1} and \eref{3.2} with those obtained from the ensemble under consideration. Then, one can use the results~\eref{3.16}, \eref{3.18} and \eref{3.19}.

All random matrix ensembles in tables~\ref{t1} and \ref{t2} have physical relevance. Real symmetric random matrices and Hermitian self-dual matrices model Hamilton operators of quantum chaotic systems \cite{Haa01}. Also circular orthogonal and circular unitary-symplectical ensembles have applications in quantum chaos. They describe Floquet operators in periodically driven systems \cite{Haa01}. Real symmetric and Hermitian self-dual chiral ensembles are used to model Dirac operators in QCD \cite{Ver94}. Furthermore one can consider all these physical systems in the presence of a chemical potential. This yields for real symmetric and Hermitian self dual matrices the corresponding Ginibre ensembles and for the chiral case the two matrix models.

The two-dimensional complex Dirac distribution used in table~\ref{t1} is defined by
\begin{equation}\label{6.1}
 \delta^{(2)}(z_1-z_2^*)=\delta(x_2-x_1)\delta(y_2+y_1)=\frac{1}{r_1}\delta(r_1-r_2)\delta(\varphi_1+\varphi_2)\,.
\end{equation}
We use the short hand notation
\begin{equation}\label{6.2}
 \eta_\pm=\frac{1\pm\mu^2}{4\mu^2}\,,
\end{equation}
c.f. Ref.~\cite{APS09b}. The functions ${\rm erfc}$ and $K_\nu$ are the complementary error--function and the $K$--Bessel function of order $\nu$, respectively. The function $f$ is calculated in Ref.~\cite{APS09b} and given by
\begin{equation}\label{6.3}
 \fl f(x+\imath y)=2\int\limits_0^\infty\exp\left[-2t(x^2-y^2)-\frac{1}{4t}\right]K_{\nu/2}(2t(x^2+y^2)){\rm erfc}(2\sqrt{t}|y|)\frac{dt}{t}\,.
\end{equation}
As in Sec.~\ref{sec4.1}, we notice that the Pfaffian structure arising in all those ensembles is fundamental. Particularly, on the formal level, the obvious difference between ensembles with orthogonal symmetry and those with unitary-symplectic symmetry becomes immaterial in our derivation. The Pfaffian structure is exclusively due to the starting point Eqs.~\eref{3.1} or \eref{3.2}. Furthermore, we expect that the list of those ensembles given here is not complete yet.

\section{Remarks and conclusions}\label{sec6}

We extended our method \cite{KieGuh09} to integrals of the types \eref{3.1} and \eref{3.2} which are averages of characteristic polynomial ratios. Those integrals have a Pfaffian structure whose kernels are averages over one or two characteristic polynomials. This coincides with known results for particular matrix ensembles \cite{Meh04,GGK04,Ake05,BorStr05} and generalizes those to the real Ginibre ensemble and the Gaussian real chiral ensemble. Tables~\ref{t1} and \ref{t2} show a wide class of ensembles for which our results are valid. Remarkably, the Pfaffian structure arising for ensembles with a real structure as well as with a quaternionic structure emerges from the same type of integral. Thus, there is no difference between both symmetries from this formal point of view.

We showed that the Pfaffian structure is a purely algebraic property for factorizing probability densities. This is a crucial new insight. It shows that supersymmetric structures appearing in the integrand can be found without mapping onto superspace. Furthermore, they are the ultimate reason for the existence of the Dyson-Mehta-Mahoux integration theorem \cite{MehMah91}. Our method also explains why one has found Pfaffian structures in so many areas of random matrix theory. As we only perform algebraic manipulations, we expect that these structures are even more general. This is indeed confirmed by the examples showed in the tables~\ref{t1} and \ref{t2}.

Surprisingly, for the case of a large number of characteristic polynomials in the denominator, the kernels reduce to one and two dimensional integrals. These integrals are the mean value of one or two characteristic polynomials in the denominator over one or two dimensional matrices, respectively. Thus in this case, we have drastically reduced the number of integrals, even below the number that would result when mapping onto superspace \cite{VWZ85,KGG08,KSG09}. This new insight shows that there are two regimes depending on the number of characteristic polynomials. In the case $k_2-k_1+2N+1>0$ the matrix in the Pfaffian determinant is up to few entries fully occupied whereas for $k_2-k_1+2N+1\leq0$ we have to take a Pfaffian determinant of a sparsely occupied matrix.

In short, Pfaffian determinants stem in our method from purely algebraic manipulations. This is the reason why our results are so general. No integration has to be performed. The Pfaffian structures are already contained in the initial integrand.

\section*{Acknowledgements}
We thank  G. Akemann, M.J. Phillips and H.-J. Sommers for fruitful discussions and for pointing out important references to us. We acknowledge support from Deutsche Forschungsgemeinschaft within Sonderforschungsbereich Transregio 12 ``Symmetries and Universality in Mesoscopic Systems''.

\appendix

\section{Details of the calculations}\label{app1}

In \ref{app1.1}, we carry out the integrals in Eq.~\eref{3.3} for the case $k_2+2N+1\geq k_1$. We derive the other case $k_2+2N+1\leq k_1$ in \ref{app1.2}.

\subsection{The case $k_2+2N+1\geq k_1$}\label{app1.1}

Let $d=k_2-k_1+2N+1\geq0$ be odd. We are interested in the integral
\begin{eqnarray}
 \fl&&\displaystyle\int\limits_{\mathbb{C}^{2N+1}}h(z_{2N+1})\prod\limits_{j=1}^{N}g(z_{2j-1},z_{2j})\sqrt{\Ber_{(k_1/k_2+2N+1)}^{(2)}(\tilde{z})}d[z]=(-1)^{k_2(k_2-1)/2+N}\times\label{a1.1}\\
 \fl&\times&\hspace*{-0.3cm}\displaystyle\int\limits_{\mathbb{C}^{2N+1}}h(z_{2N+1})\prod\limits_{j=1}^{N}g(z_{2j-1},z_{2j})\det\left[\begin{array}{cc} \left\{\displaystyle\frac{1}{\kappa_{a1}-\kappa_{b2}}\right\}\underset{1\leq b\leq k_2}{\underset{1\leq a\leq k_1}{\ }} & \left\{\displaystyle\frac{1}{\kappa_{a1}-z_{b}}\right\}\underset{1\leq b\leq 2N+1}{\underset{1\leq a\leq k_1}{\ }} \\ \left\{\displaystyle\kappa_{b2}^{a-1}\right\}\underset{1\leq b\leq k_2}{\underset{1\leq a\leq d}{\ }} & \left\{\displaystyle z_b^{a-1}\right\}\underset{1\leq b\leq 2N+1}{\underset{1\leq a\leq d}{\ }} \end{array}\right]d[z]\,.\nonumber
\end{eqnarray}
The first step is the integration over all variables $z_j$ with an odd index $j$. Thus, we have
\begin{eqnarray}
 \fl&&\displaystyle\int\limits_{\mathbb{C}^{2N+1}}h(z_{2N+1})\prod\limits_{j=1}^{N}g(z_{2j-1},z_{2j})\sqrt{\Ber_{(k_1/2N+1+k_2)}^{(2)}(\tilde{z})}d[z]=(-1)^{k_2(k_2-1)/2+N}\times\label{a1.2}\\
 \fl&\times&\displaystyle\int\limits_{\mathbb{C}^{N}}\det\left[\begin{array}{cc} \left\{\displaystyle\frac{1}{\kappa_{b1}-\kappa_{a2}}\right\}\underset{1\leq b\leq k_1}{\underset{1\leq a\leq k_2}{\ }} & \left\{\displaystyle\kappa_{a2}^{b-1}\right\}\underset{1\leq b\leq d}{\underset{1\leq a\leq k_2}{\ }} \\ \left\{\displaystyle\int\limits_{\mathbb{C}}\frac{h(z)}{\kappa_{b1}-z}dz\right\}\underset{1\leq b\leq k_1}{\ } & \left\{\displaystyle\int\limits_{\mathbb{C}}h(z)z^{b-1}dz\right\}\underset{1\leq b\leq d}{\ } \\ \left\{\begin{array}{c} \displaystyle\int\limits_{\mathbb{C}}\frac{g(z,z_a)}{\kappa_{b1}-z}dz \\ \displaystyle \frac{1}{\kappa_{b1}-z_{a}}\end{array}\right\}\underset{1\leq b\leq k_1}{\underset{1\leq a\leq N}{\ }} & \left\{\begin{array}{c}\displaystyle\int\limits_{\mathbb{C}}g(z,z_a)z^{b-1}dz \\ \displaystyle z_a^{b-1} \end{array}\right\}\underset{1\leq b\leq d}{\underset{1\leq a\leq N}{\ }} \end{array}\right]d[z]\,.\nonumber
\end{eqnarray}
We perform the last integrals with help  of a modified de Bruijn's integral theorem \cite{Bru55,KieGuh09} and find
\begin{eqnarray}
 \fl&&\displaystyle\int\limits_{\mathbb{C}^{2N+1}}h(z_{2N+1})\prod\limits_{j=1}^{N}g(z_{2j-1},z_{2j})\sqrt{\Ber_{(k_1/2N+1+k_2)}^{(2)}(\tilde{z})}d[z]=(-1)^{N+1}N!\times\nonumber\\
 \fl&\times&\displaystyle\Pf\left[\begin{array}{ccc} 0 & \left\{\displaystyle\frac{1}{\kappa_{b1}-\kappa_{a2}}\right\}\underset{1\leq b\leq k_1}{\underset{1\leq a\leq k_2}{\ }} & \left\{\mathbf{K}_{(d)}(\kappa_{a2})\right\}\underset{1\leq a\leq k_2}{\ }  \\ \left\{-\displaystyle\frac{1}{\kappa_{a1}-\kappa_{b2}}\right\}\underset{1\leq b\leq k_2}{\underset{1\leq a\leq k_1}{\ }} & \left\{\mathbf{F}(\kappa_{a1},\kappa_{b1})\right\}_{1\leq a,b \leq k_1} & \left\{\mathbf{G}_{(d)}(\kappa_{a1})\right\}_{1\leq a\leq k_1} \\ \left\{-\mathbf{K}_{(d)}^T(\kappa_{b2})\right\}\underset{1\leq b\leq k_2}{\ } & \left\{-\mathbf{G}_{(d)}^T(\kappa_{b1})\right\}_{1\leq b\leq k_1} & \mathbf{M}_{(d)} \end{array}\right]\label{a1.3}
\end{eqnarray}
with the matrices defined in Eqs. (\ref{3.5}-\ref{3.11}). Finally, we extract the matrix $\mathbf{M}_{(d)}$ from the Pfaffian by inversion, see Eq.~\eref{2.10}, and arrive at Eq.~\eref{3.4}.

\subsection{The case $k_2+2N+1\leq k_1$}\label{app1.2}

Let $d=k_2-k_1+2N+1\leq0$ be an arbitrary integer. Then, we calculate
\begin{eqnarray}
 \fl&&\displaystyle\int\limits_{\mathbb{C}^{2N+1}}h(z_{2N+1})\prod\limits_{j=1}^{N}g(z_{2j-1},z_{2j})\sqrt{\Ber_{(k_1/k_2+2N+1)}^{(2)}(\tilde{z})}d[z]=\nonumber\\
 \fl&=&(-1)^{k_1(k_1-1)/2+k_1-k_2-1}\displaystyle\int\limits_{\mathbb{C}^{2N+1}}h(z_{2N+1})\prod\limits_{j=1}^{N}g(z_{2j-1},z_{2j})\times\nonumber\\
 \fl&\times&\det\left[\begin{array}{ccc} \left\{\displaystyle\frac{1}{\kappa_{a1}-\kappa_{b2}}\right\}\underset{1\leq b\leq k_2}{\underset{1\leq a\leq k_1}{\ }} & \left\{\displaystyle \kappa_{a1}^{b-1}\right\}\underset{1\leq b\leq -d}{\underset{1\leq a\leq k_1}{\ }} & \left\{\displaystyle\frac{1}{\kappa_{a1}-z_{b}}\right\}\underset{1\leq b\leq 2N+1}{\underset{1\leq a\leq k_1}{\ }} \end{array}\right]d[z]\,.\label{a1.4}
\end{eqnarray}
As in \ref{app1.1}, we integrate first over all variables with an odd index. This yields
\begin{eqnarray}
 \fl&&\displaystyle\int\limits_{\mathbb{C}^{2N+1}}h(z_{2N+1})\prod\limits_{j=1}^{N}g(z_{2j-1},z_{2j})\sqrt{\Ber_{(k_1/k_2+2N+1)}^{(2)}(\tilde{z})}d[z]=\nonumber\\
 \fl&=&(-1)^{k_1(k_1-1)/2+k_1-k_2-1}\displaystyle\int\limits_{\mathbb{C}^{N}}\det\left[\begin{array}{c} \left\{\displaystyle\frac{1}{\kappa_{b1}-\kappa_{a2}}\right\}\underset{1\leq b\leq k_1}{\underset{1\leq a\leq k_2}{\ }} \\ \left\{\displaystyle \kappa_{b1}^{a-1}\right\}\underset{1\leq b\leq k_1}{\underset{1\leq a\leq -d}{\ }} \\ \left\{\displaystyle\int\limits_{\mathbb{C}}\frac{h(z)}{\kappa_{b1}-z}d[z]\right\}_{1\leq b\leq k_1} \\ \left\{\begin{array}{c}\displaystyle\int\limits_{\mathbb{C}}\frac{g(z,z_b)}{\kappa_{b1}-z}d[z] \\ \displaystyle\frac{1}{\kappa_{b1}-z} \end{array}\right\}\underset{1\leq b\leq k_1}{\underset{1\leq a\leq N}{\ }} \end{array}\right]d[z]\,.\label{a1.5}
\end{eqnarray}
Again we use the modified version of de Bruijn's integral theorem and obtain Eq.~\eref{3.19} up to the Berezinian in the denominator.

\section*{References}


\end{document}